\documentclass{article}
\usepackage{spconf,amsmath,amsfonts,graphicx}
\usepackage{algorithmic}
\usepackage{array}
\usepackage{url}
\usepackage{graphicx}
\usepackage{subfigure} 
\usepackage{booktabs}
\usepackage{multirow}
\usepackage{makecell}
\usepackage{bm}
\usepackage{enumitem}
\usepackage{threeparttable}

\title{Target-Speaker Voice Activity Detection via Sequence-to-Sequence Prediction}
\name{Ming~Cheng$^{1,2}$, Weiqing~Wang$^{2}$, Yucong~Zhang$^{2}$, Xiaoyi~Qin$^{1,2}$, Ming~Li$^{1,2\dagger}$\thanks{$\dagger$ Corresponding Author, E-mail: ming.li369@dukekunshan.edu.cn}}

\address{
$^1$School of Computer Science, Wuhan University, Wuhan, China\\
$^2$Data Science Research Center, Duke Kunshan University, Kunshan, China\\
}

\begin{document}
\ninept

\maketitle

\begin{abstract}
Target-speaker voice activity detection is currently a promising approach for speaker diarization in complex acoustic environments. This paper presents a novel Sequence-to-Sequence Target-Speaker Voice Activity Detection (Seq2Seq-TSVAD) method that can efficiently address the joint modeling of large-scale speakers and predict high-resolution voice activities. Experimental results show that larger speaker capacity and higher output resolution can significantly reduce the diarization error rate (DER), which achieves the new state-of-the-art performance of 4.55\% on the VoxConverse test set and 10.77\% on Track 1 of the DIHARD-III evaluation set under the widely-used evaluation metrics.
\end{abstract}

\section{Introduction}
Speaker diarization refers to the task of determining each speaker's utterance boundaries in multi-party conversational audio~\cite{tranter2006overview}. As a front-end processing to segment audio into pieces with different identities, speaker diarization can effectively participate in various downstream tasks~\cite{park2022review}.

Conventional speaker diarization consists of several independent modules. First, Voice Activity Detection (VAD)~\cite{chang2018temporal} removes non-speech regions from the input audio. Then, the remaining is partitioned into multiple short segments, followed by speaker embedding extraction~\cite{snyder2018x}. By measuring segment-wise speaker similarities with cosine distance or PLDA~\cite{prince2007probabilistic}, clustering-based methods can group segmented utterances into different identities~\cite{lin2019lstm}.

One limitation of conventional approaches is that they cannot handle overlapped speech, as each audio segment is assumed to contain only one speaker. To tackle this problem, target-speaker voice activity detection (TS-VAD)~\cite{medennikov2020target} has attracted much interest due to its great success in challenging tasks such as VoxSRC~\cite{wang2021dku,wang2022dku}, and DIHARD-III~\cite{wang2021ustc}. Based on speaker profiles from a clustering-based diarization, the TS-VAD system can estimate each speaker's frame-level voice activities to refine the initial clustering-based results.

\begin{figure}[t]
\centering
  \includegraphics[width=0.9\linewidth]{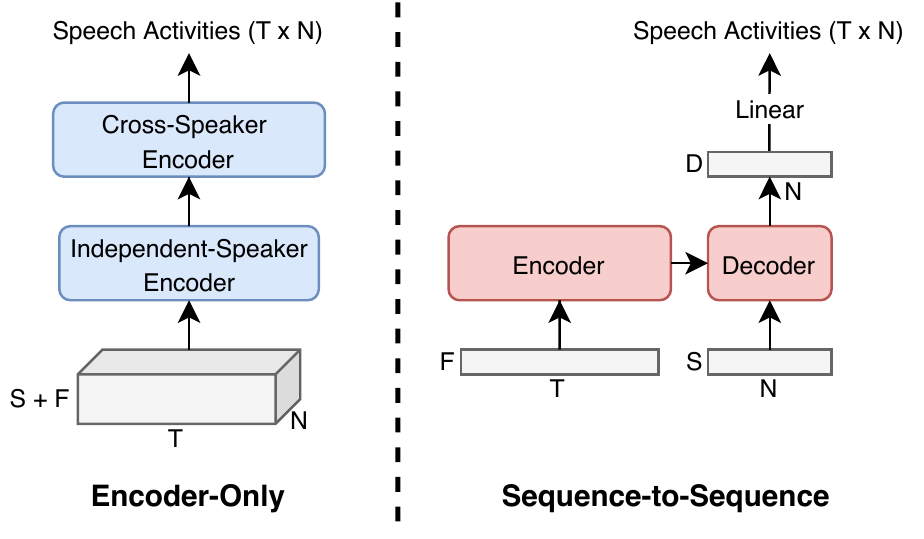}
  \caption{Diagram of encoder-only TS-VAD method (left) and our proposed sequence-to-sequence method (right). $T$ and $F$ denote the length and dimension of extracted frame-level features. $N$ and $S$ denote the number and dimension of speaker embeddings. $D$ represents the output dimension of the decoder.}
  \label{fig:intro}
\end{figure}

In recent years, many researchers improve the TS-VAD system by modifications on front-end preprocess~\cite{wang2021scenario}, independent-speaker prediction~\cite{9849033}, and cross-speaker~\cite{wang2022target} modeling. We abstract the usual TS-VAD method into the left part of Fig.~\ref{fig:intro}, namely the encoder-only method. The front-end extractor is omitted in the plot, which extracts frame-level representations $\mathbf{E_{rep}} \in \mathbb{R}^\mathrm{T \times F}$ from the input audio. Let $\mathbf{E_{spk}} \in \mathbb{R}^\mathrm{N \times S}$ denote the given speaker profiles. Each target-speaker embedding in $\mathbf{E_{spk}}$ has to be repeated $T$ times and concatenated with the $\mathbf{E_{rep}}$ to produce a 3-dimensional input tensor with the shape of $T \times N \times (S+F)$. Then, various encoders (e.g., Bi-LSTM, Self-Attention) predict speech activities by processing the input data along the time-axis ($T$) and speaker-axis ($N$). However, the encoder-only method leads to high demand for GPU memory since the space consumption of the input tensor is proportional to $T \times N \times (S+F)$. With the growth of $T$ and $N$, the dramatic boost of memory usage limits the model to process longer feature sequences and more speakers at once. Furthermore, the output length of encoder-only models must be equal to $T$, which means the temporal resolution of voice activity detection (VAD) cannot be changed arbitrarily.

Considering the limitations above, we propose a sequence-to-sequence framework for target-speaker voice activity detection, shown in the right part of Fig.~\ref{fig:intro}. Our main contributions include:

\begin{figure*}[t]
\centering
	\subfigure[Overview]{
	\begin{minipage}[b]{0.65\linewidth}
		\centering
		\label{fig:framework}
		\includegraphics[height=6cm]{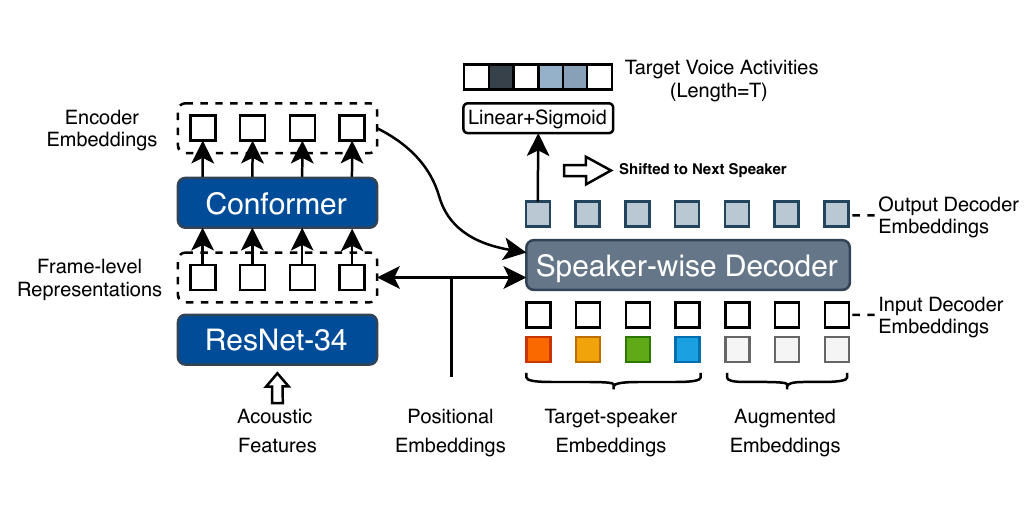}
		\end{minipage}%
		}%
	\subfigure[Speaker-wise Decoder]{
	\begin{minipage}[b]{0.4\linewidth}
		\centering
		\label{fig:decoder}
		\includegraphics[height=6cm]{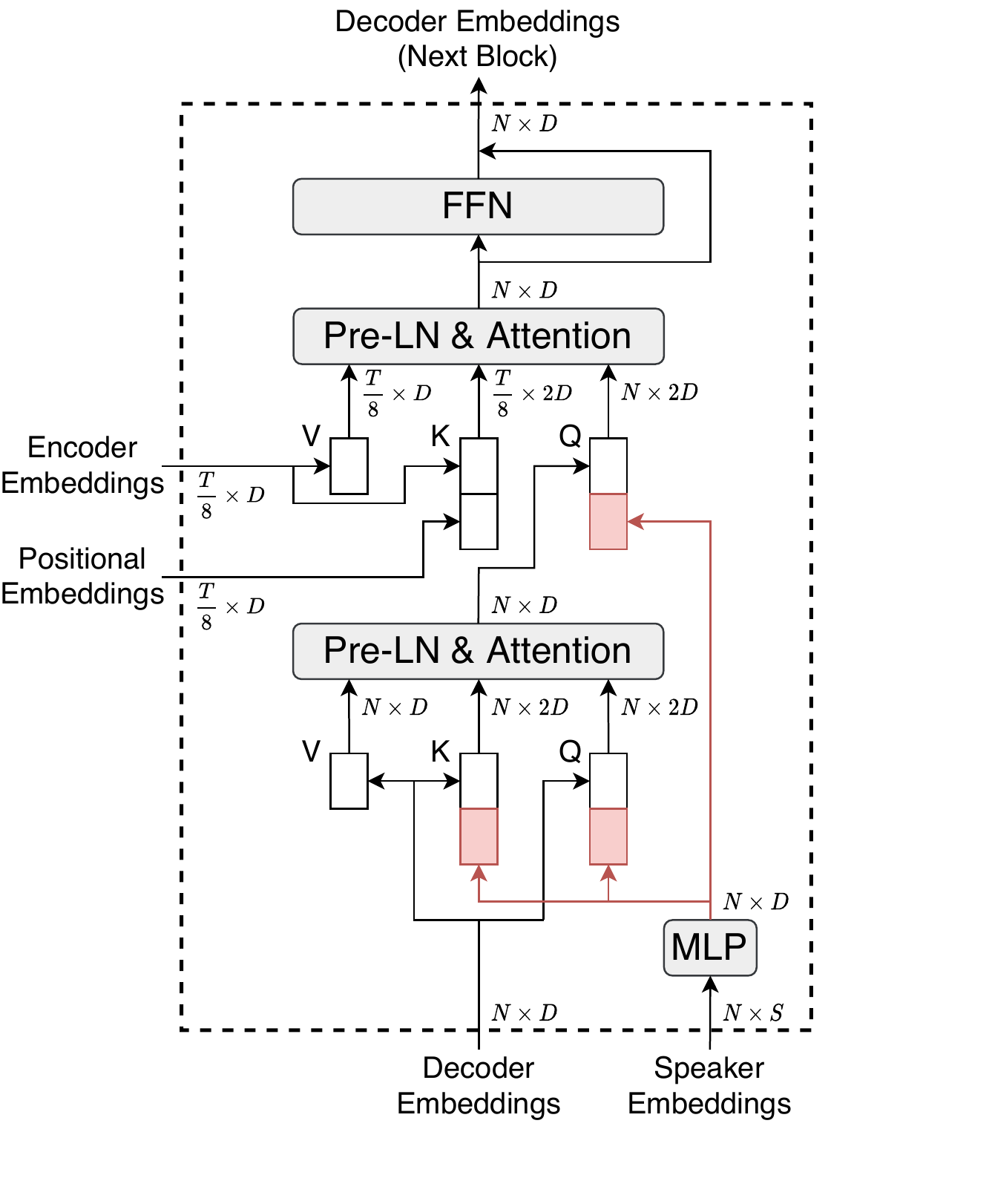}
		\end{minipage}%
	}%
	\caption{The Sequence-to-Sequence Target-Speaker Voice Activity Detection (Seq2Seq-TSVAD) Framework.}
	\label{fig:overall}
\end{figure*}

\begin{enumerate}
\setlength{\itemsep}{0pt}
	\item Frame-level representations and speaker embeddings are separately fed into the encoder and decoder sides. Thus, the memory consumption of the inputs becomes proportional to $T \times F + N \times S$, which reduces redundant tensors.
	\item The decoder compresses each speaker's voice activity information into a fixed-dim ($D$) embedding regardless of the length of input features. The last linear layer can adjust the model to predict voice activities in higher temporal resolution with a small computational cost.
	\item Experimental results show that joint modeling of more speakers with higher VAD resolution can reduce the diarization error rate (DER) significantly, which achieves the new state-of-the-art DERs of 4.55\% on the VoxConverse~\cite{chung2020spot} test set and 10.77\% on Track 1 of the DIHARD-III~\cite{ryant2020third} evaluation set.
\end{enumerate}

\section{Methods}

The framework of our proposed Seq2Seq-TSVAD method is shown in Fig.~\ref{fig:framework}. The details are introduced as follows.

\subsection{ResNet-34}
The front-end feature extractor is ResNet-34~\cite{he2016deep}, where the widths (number of channels) of the residual blocks are $\left\{64,128,256,512\right\}$. Given $\mathbf{X} \in \mathbb{R}^\mathrm{T \times H}$ denoting the acoustic feature sequence with the length of $T$ and dimension of $H$, the model outputs a CNN feature map $\mathbf{M} \in \mathbb{R}^\mathrm{C \times \frac{T}{8}  \times \frac{H}{8} }$, where $C$ is the number of channels. Then, we adopt the segmental statistical pooling~\cite{9849033} (SSP) method to aggregate the channel-wise feature map into frame-level representations $\mathbf{E_{rep}} \in \mathbb{R}^\mathrm{\frac{T}{8}  \times {F}}$, where ${F}$ denotes the output dimension of the SSP layer. This process can be viewed as the downsampling from the input acoustic feature sequence.

\subsection{Conformer}
We introduce the Conformer~\cite{gulati2020conformer} to further model long-term dependencies between frame-level representations. The input layer of the Conformer module firstly maps $\mathbf{E_{rep}}$ to $\mathbf{{E_{rep}}'} \in \mathbb{R}^\mathrm{\frac{T}{8}  \times {D}}$ with sinusoidal positional encodings, where $D$ is equal to the attention size. Then, the input sequence $\mathbf{{E_{rep}}'}$ is encoded into $\mathbf{E_{enc}} \in \mathbb{R}^\mathrm{\frac{T}{8}  \times {D}}$ by the following Conformer blocks.

\subsection{Speaker-wise Decoder}

The Speaker-wise Decoder (SW-D) estimates target-speaker voice activities by considering cross-speaker correlations. The input on the decoder side can be broken into two parts: decoder embeddings and auxiliary queries. In this case, the input decoder embeddings are initialized by zeros, and target-speaker embeddings are fed as auxiliary queries. 

Let $\mathbf{E_{spk}} = \left [\mathbf{e_{1}} \ldots \mathbf{e_{N}} \right ]^{\mathrm{T}} \in \mathbb{R}^\mathrm{N  \times S}$ denote the given target-speaker embeddings, where $N$ is the number of speakers and $S$ is the embedding dimension. The SW-D module incorporates $\mathbf{E_{enc}}$ and $\mathbf{E_{spk}}$ to output processed decoder embeddings which contain information about each speaker's voice activities. Let $\mathbf{E_{dec}} = \left [\mathbf{v_{1}} \ldots \mathbf{v_{N}} \right ]^{\mathrm{T}} \in \mathbb{R}^\mathrm{N \times D}$ denote the predicted decoder embeddings, where $D$ is equal to the attention size used in Conformer blocks. After a linear projection with sigmoid activation, each $\mathbf{v_{i}}$ can be transformed into posterior probabilities of voice activities belonging to the \textit{i-th} speaker. The output dimension of the last linear layer controls the temporal resolution of detected voice activities, which means a longer output length can provide more precise timestamps.

Fig.~\ref{fig:decoder} demonstrates the structure of the SW-D module. Inspired by DAB-DETR~\cite{liu2022dab}, we modify the vanilla Transformer decoder so that speaker embeddings can be introduced into the attention mechanism. In each decoder block, target-speaker embeddings go through a multi-layer perception (MLP) to generate the within-block representations. Before the first attention layer, the within-block embeddings are concatenated with the original keys and queries to calculate attention scores. Since decoder embeddings are designed to learn positional (temporal) information about voice activities, we also concatenate keys from encoders with corresponding positional embeddings. This way, all key-query calculations can incorporate positional and speaker-related information. The decoding process effectively exchanges information between all speakers by self-attention mechanism.

The adopted MLP module consists of two linear layers with in-between Layer Normalization and ReLU activation. The Pre-LayerNorm method~\cite{xiong2020layer} is employed before each attention layer. For clarity, we omit to plot residual connections of attention layers.

\subsection{Embedding Augmentation}

To align the different speaker numbers in training mini-batch data, a set of augmented embeddings for each audio sample is supplied to ensure that the SW-D module has a fixed decoding length. Each augmented embedding may be directly set to zeros with a probability of 0.5. Otherwise, it will be padded by another one not appearing in the current audio. Then, there is a probability of 0.2 to replace all input speaker embeddings with non-existent speakers. Furthermore, the input embeddings and corresponding ground-truth labels should be shuffled to keep the model invariant to speaker order. This augmentation method is vital because it forces the network to learn to distinguish valid or invalid speakers in given audio.

\section{Experimental Results}

\subsection{Speaker Embedding}
We adopt the ResNet-34 model as the pattern extractor. After the statistical pooling~\cite{snyder2018x} layer, a linear projection outputs the 256-dim speaker embedding. The ArcFace (s=32,m=0.2)~\cite{deng2019arcface} is used as the classifier. The detailed configuration of the neural network is the same as~\cite{wang2020dku}. The model trained on the VoxCeleb2~\cite{Chung18b} dataset finally achieves an equal error rate (EER) of 0.814\% on the Vox-O~\cite{Nagrani17} trial. The pre-trained model is used for speaker embedding extraction in the subsequent Seq2Seq-TSVAD system.

\subsection{Seq2Seq-TSVAD Configuration}
\label{tsvad_config}

\subsubsection{Model Parameters}

The front-end ResNet-34 is initialized by the pre-trained speaker embedding model. We replace the original statistical pooling (SP) with a segmental statistical pooling (SSP)~\cite{9849033} layer to extract frame-level features. The back-end model consists of Conformer encoders~\cite{gulati2020conformer} and Transformer decoders~\cite{vaswani2017attention}. All encoder-decoder modules have 6 blocks sharing the same settings: 512-dim attentions with 8 heads and 1024-dim feed-forward layers with a dropout rate of 0.1. The kernel size of convolutions in Conformer blocks is 15, and the other implementation details are the same as~\cite{gulati2020conformer}.

\subsubsection{Data Simulation}
As neural network-based methods can usually benefit from large-scale training data, we create a simulated corpus by the VoxCeleb2~\cite{Chung18b} and LibriSpeech~\cite{7178964} datasets. The WebRTC\footnote{\url{https://github.com/wiseman/py-webrtcvad}} toolkit is used to remove non-speech regions from raw audio. Then, each simulated signal is generated by mixing source utterances of 1-4 speakers from the processed corpus. The overlap ratio of simulated data has not been controlled because each single-speaker component is produced randomly and individually in an online process.

\subsubsection{Training Process}
All training audio signals are split into 16-sec chunks. The model input takes 80-dim log Mel-filterbank energies with a frame length of 25 ms and a frameshift of 10 ms as acoustic features. Furthermore, background noise from Musan~\cite{snyder2015musan} and reverberation from RIRs~\cite{7953152} are applied as the data augmentation.
 
We implement the BCE loss and Adam optimizer to train the neural network with a linear learning rate warm-up. Firstly, the model with frozen ResNet-34 is trained by simulated data until back-end convergence. Then, all model parameters are unfrozen to keep them trainable. The real data from the specific dataset is added to the simulated data at a ratio of 0.2. Finally, we finetune the model on real data without any simulation. The first two stages take around 200 epochs with a learning rate of \textit{1e-4}, and the last stage decreases the learning rate to \textit{1e-5}.

\subsubsection{Inference Settings}

A clustering-based diarization first provides an initial result in the inference stage. Based on that, the pre-trained speaker embedding model extracts target embeddings from speech segments. The speakers with non-overlapped speech shorter than 2 seconds are discarded. 

We split each test recording into 16-sec chunks and feed them into the TS-VAD  model with extracted speaker embeddings. Then, the final predictions are stitched chunk by chunk. The maximum number of speaker profiles (decoding length) is set to $L$. If the detected speakers are less than $L$, the rest of the speaker embeddings will be padded by zeros. Otherwise, the extra part will be inferred in the next group.

Lastly, we adopt the general voice activity detection (VAD) to revise the TS-VAD predictions as post-processing. According to specific evaluation requirements, Oracle or Estimated VAD information can be obtained from the initial diarization result. The timestamps marked as active speech will directly assign a positive label to the speaker with the highest predicted score. Predictions at the timestamps marked as non-speech will be zeroed.

\begin{table}[t]
	\centering
	\setlength{\tabcolsep}{6.18pt}
	\renewcommand{\arraystretch}{0.8}
	\caption{DERs (\%) of different Seq2Seq-TSVAD models on the VoxConverse test set (collar = 250 ms). $\ast$ denotes the model training without embedding augmentation.}

	\label{voxconverse_exps}
	\begin{tabular}{llrrr}
		\toprule
		\multirow{2}{*}{\textbf{\makecell[c]{VAD\\Resolution}}} & \multirow{2}{*}{\textbf{\makecell[c]{Decoding\\Length}}} & \multicolumn{3}{c}{\textbf{DER (\%)}} \\
		\cmidrule(lr){3-5} 
		& & \textbf{1-10 SPKs} & \textbf{10+ SPKs} & \textbf{Total} \\
		\midrule
		\multirow{3}{*}{R = 80 ms} 
		& L = 10 & 3.94 & 9.16 &  5.01\\ 
		& L = 20 & 4.03 & 7.08 & 4.66 \\
		& L = 30 & 3.96 & 7.03 & \textbf{4.59} \\
		& L = 30 $^\ast$ & 9.07 & 21.66 & 11.65\\
		\midrule
		\multirow{3}{*}{R = 10 ms} 
		& L = 10 & 4.03 & 10.29 & 5.32 \\ 
		& L = 20 & 3.99 & 7.10  & 4.63\\
		& L = 30 & 3.94 & 6.93  & \textbf{4.55} \\
		& L = 30 $^\ast$ & 8.45 & 16.52 & 10.10\\
		\bottomrule
	\end{tabular}
\end{table}

\begin{table}[t]
	\centering
	\setlength{\tabcolsep}{3.5pt}
	\renewcommand{\arraystretch}{0.8}
	\caption{Comparisons of our proposed Seq2Seq-TSVAD models with others on the VoxConverse test set (collar = 250 ms).}
	\label{voxconverse_comp}
	\begin{threeparttable}[b]
	\begin{tabular}{lrr}
		\toprule
		\textbf{Method} & \textbf{DER (\%)} & \textbf{JER (\%)} \\
		\midrule
		ByteDance~\cite{wang2021bytedance} \tnote{$\dag$} & 5.17 & 29.56 \\
		DKU-DukeECE~\cite{wang2022dku} \tnote{$\ddag$} & 4.94 & 28.79 \\
		Wang et al.~\cite{wang2022target}  \tnote{$\ast$} & 4.57 & - \\  
		\midrule
		AHC~\cite{wang2022dku} & 5.35 & 27.99 \\
		\quad $+$ Seq2Seq-TSVAD (R=80ms, L=30) & 4.59 & 29.39 \\
		\quad $+$ Seq2Seq-TSVAD (R=10ms, L=30) & \textbf{4.55} & 29.13 \\
		\bottomrule
	\end{tabular}
	\begin{tablenotes}
    	\item[$\dag$]  VoxSRC-21 2nd-ranked result with 3-system fusion. The top-ranked team did not report the DER/JER for this set. 
    	\item[$\ddag$] VoxSRC-22 1st-ranked result with 4-system fusion.
    	\item[$\ast$] The authors update this result from 4.74\% to 4.57\% in their latest version.
    \end{tablenotes}
	\end{threeparttable}
\end{table}

\subsection{Evaluation}

\subsubsection{VoxConverse Dataset}

We directly copy our AHC-based diarization model used in VoxSRC-22~\cite{wang2022dku} to be the initial system for speaker embedding extraction, which obtains a DER of 5.35\% on the VoxConverse~\cite{chung2020spot} test set with a collar of 250 ms. 

According to the instructions described in Section~\ref{tsvad_config}, we train and evaluate our proposed Seq2Seq-TSVAD models using different VAD resolutions (duration per frame-level prediction) and decoding lengths (maximum capacity for speaker embeddings). Two types of VAD resolutions are provided as coarse (80 ms) and precise (10 ms) options, which can be implemented easily by adjusting the dimension of the output layer in our model. As the VoxConverse test set has a maximum of 21 speakers in a single recording, we choose the decoding lengths of 10, 20, and 30 to cover situations of insufficient, suitable, and sufficient speaker capacities, respectively. 

Table~\ref{voxconverse_exps} shows that the total DER reduces obviously when the decoding length increases from 10 to 20. And the performance gain mainly comes from the subset of recordings with over 10 speakers. In addition, a longer decoding length of 30 still brings a slight improvement. It can be seen that the precise VAD resolution (10 ms) achieves the lowest DER of 4.55\%. Table~\ref{voxconverse_comp} compares our proposed method with the current state-of-the-art results. Our best performance significantly outperforms previous ones, especially for some multi-system fusion results in past VoxSRC challenges.

\subsubsection{DIHARD-III Dataset}

\begin{table}[t]
	\centering
	\setlength{\tabcolsep}{7.0pt}
	\renewcommand{\arraystretch}{0.8}
	\caption{DERs (\%) of different Seq2Seq-TSVAD models on Track 1 of the DIHARD-III evaluation set (Oracle VAD). $\ast$ denotes the model training without embedding augmentation.}
	\label{dihard_exps}
	\begin{tabular}{llrrr}
		\toprule
		\multirow{2}{*}{\textbf{\makecell[c]{VAD\\Resolution}}} & \multirow{2}{*}{\textbf{\makecell[c]{Decoding\\Length}}} & \multicolumn{3}{c}{\textbf{DER (\%)}} \\
		\cmidrule(lr){3-5} 
		& & \textbf{1-5 SPKs} & \textbf{5+ SPKs} & \textbf{Total} \\
		\midrule
		\multirow{3}{*}{R = 80 ms} 
		& L = 5  & 10.46 & 24.57 & 12.44 \\ 
		& L = 10 & 10.56 & 23.00 & 12.30 \\
		& L = 20 & 10.42 & 22.69 & \textbf{12.14} \\
		& L = 20 $^\ast$ & 10.73 & 25.40 & 12.79 \\
		\midrule
		\multirow{3}{*}{R = 10 ms} 
		& L = 5  & 8.91 & 23.96 & 11.02 \\ 
		& L = 10 & 8.97 & 22.02 & 10.80 \\
		& L = 20 & 8.94 & 21.99 & \textbf{10.77} \\
		& L = 20 $^\ast$ & 9.29 & 24.25 & 11.39 \\
		\bottomrule
	\end{tabular}
\end{table}

\begin{table}[t]
	\centering
	\setlength{\tabcolsep}{3.3pt}
	\renewcommand{\arraystretch}{0.8}
	\caption{Comparisons of our proposed Seq2Seq-TSVAD models with others on Track 1 of the DIHARD-III evaluation set (Oracle VAD).}
	\label{dihard_comp}
	\begin{threeparttable}[b]
	\begin{tabular}{lrr}
		\toprule
		\textbf{Method} & \textbf{DER (\%)} & \textbf{JER (\%)}  \\
		\midrule
		USTC-NELSLIP~\cite{wang2021ustc} \tnote{$\dag$}  & 11.30 & 29.94 \\
		Hitachi-JHU~\cite{horiguchi2021hitachi} \tnote{$\ddag$} & 11.58 & 32.37 \\
		Wang et al.~\cite{wang2021scenario} & 11.30 & -\\
		\midrule
		LSTM-SC~\cite{9849033} & 15.40 & 33.27\\
		\quad $+$ Seq2Seq-TSVAD (R=80ms, L=20) & 12.14 & 29.49 \\
		\quad $+$ Seq2Seq-TSVAD (R=10ms, L=20) & \textbf{10.77} & 28.46 \\
		\bottomrule
	\end{tabular}
	\begin{tablenotes}
    	\item[$\dag$]  DIHARD3-III 1st-ranked result with 5-system fusion.
    	\item[$\ddag$] DIHARD3-III 2nd-ranked result with 5-system fusion.
    \end{tablenotes}
    \end{threeparttable}
\end{table}

To evaluate our proposed method on highly overlapped speech with complex background noises, we also test it on Track 1 of DIHARD-III Diarization Challenge~\cite{ryant2020third}. The LSTM-SC method in~\cite{9849033} is adopted as the initial system for speaker embedding extraction, which has a DER of 15.4\% on the evaluation set with Oracle VAD.

Since there are up to 9 speakers can appear in the DIHARD-III recordings, we explore decoding lengths of 5, 10, and 20. Table~\ref{dihard_exps} illustrates that the longer decoding length and more precise VAD resolution can bring a better result, which has the same conclusion as Table~\ref{voxconverse_exps}. Since the DIHARD-III dataset is annotated at 10-ms levels and evaluated without a tolerance collar, it is sensitive to utterance boundaries. Thus, the benefits of precise VAD resolution (10 ms) on it seem more significant than VoxConverse Dataset. 

Table~\ref{dihard_comp} shows the comparisons of our proposed method with others. Our best single system obtains a DER of 10.77\% on Track 1 of the DIHARD-III evaluation set, demonstrating a competitive performance over previous state-of-the-art results.

\begin{figure}[t]
\centering
  \includegraphics[width=\linewidth]{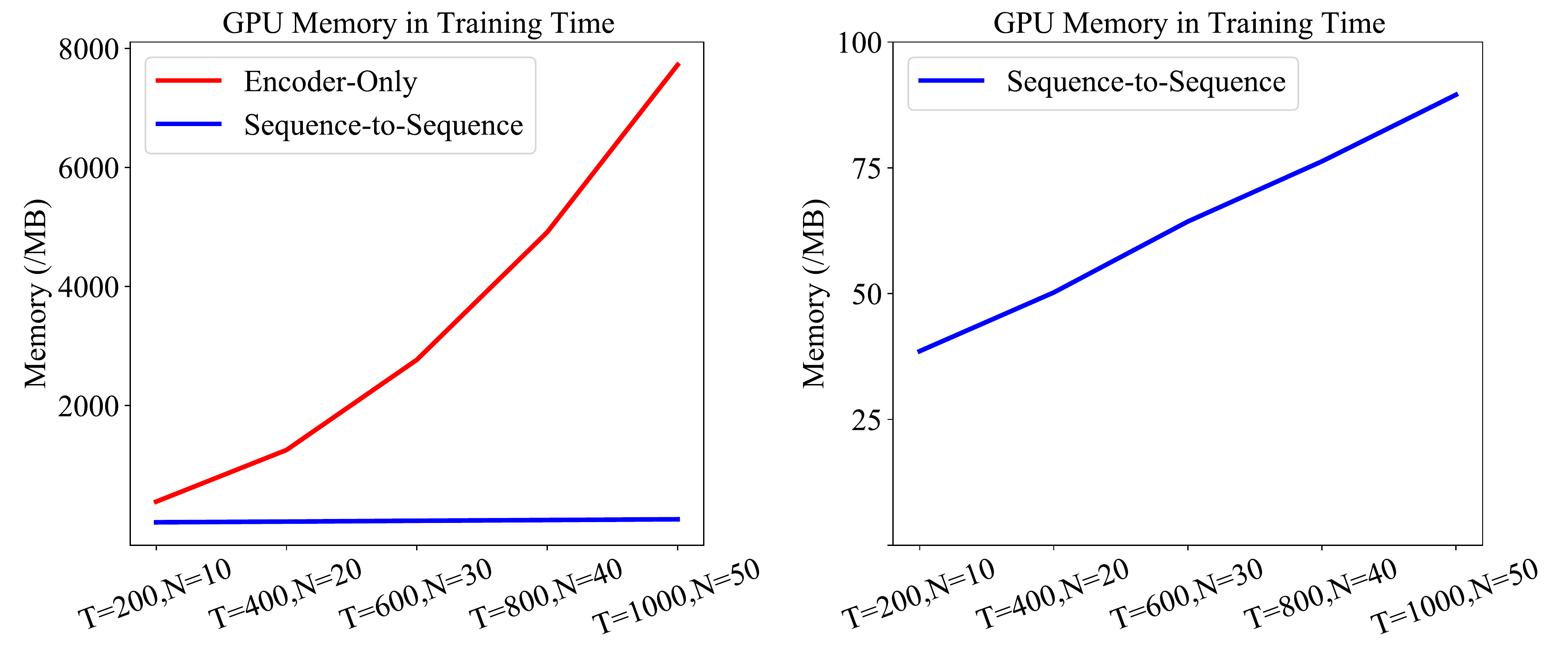}
  \caption{Visualization of overall GPU memory usage in designed training configurations.}
  \label{fig:memory}
\end{figure}

\subsection{GPU Memory Visualization}

There are usually many variable components in a TS-VAD system (e.g., front-end extractors, encoders, decoders), and it is not easy to precisely compare the memory usage of our proposed method with others. To simplify the ablation test, we only count the encoder-decoder modules shown in Fig.~\ref{fig:intro}. The encoders in the two frameworks are Conformer blocks, and the decoder in the Seq2Seq-TSVAD is our modified speaker-wise decoder. All encoder-decoder modules are set to contain a single block.

By fixing the $S$, $F$, and $D$ equal to 256 (dimension of speaker embedding), Fig.~\ref{fig:memory} visualizes the memory usage of two frameworks in different combinations of $T$ and $N$. With the increase of $T$ and $N$, the memory usage of the encoder-only TS-VAD method shows exponential growth, whereas the Seq2Seq-TSVAD method has a nearly linear trend. It should be noted that the numerical results are not absolutely precise because there are also a series of extra memory costs during model training (e.g., GPU context, optimizer, gradient checkpoints). In practice, the Seq2Seq-TSVAD does not need to adopt longer $T$ for higher VAD resolution but adjusts the last linear layer, which can save more GPU memory.

\section{Conclusions}

This paper presents a novel Sequence-to-Sequence Target-Speaker Voice Activity Detection (Seq2Seq-TSVAD) method to be scalable in terms of speaker capacity and temporal resolution. By factorizing the time-axis ($T$) and speaker-axis ($N$) of the original TS-VAD input, our proposed method reduces its space complexity from $\mathcal{O}(T \times N)$ into $\mathcal{O}(T + N)$, where $S$ and $F$ are ignored as constant values. Consequently, the novel framework can efficiently deal with 30 speakers to predict 10ms-level voice activities in a single 12-GB GPU device, which still does not reach the upper limit. Meanwhile, it obtains new state-of-the-art DERs of 4.55\% on the VoxConverse~\cite{chung2020spot} test set and 10.77\% on Track 1 of the DIHARD-III~\cite{ryant2020third} evaluation set, respectively.
 
\section{Acknowledgments}

This research is funded in part by the National Natural Science Foundation of China (62171207), Science and Technology Program of Guangzhou City (202007030011), Science and Technology Program of Suzhou City (SYC2022051). Many thanks for the computational resource provided by the Advanced Computing East China Sub-Center.

\bibliographystyle{IEEE}
\bibliography{strings,refs}

\end{document}